\documentclass{sig-alternate}
\usepackage{algorithm,algpseudocode,cite,amsmath,amssymb,amsfonts,dsfont,multirow,multicol,enumitem,epsfig,url,array,makecell,balance}
\usepackage{times}

\newtheorem{definition} {Definition}
\newtheorem{example} {Example}
\newtheorem{theorem} {Theorem}

\newtheorem{observation} {Observation}

\def\done{\hspace*{\fill} {$\square$}}

\def\header{\vspace{2mm} \noindent}

\def\figcapup{\vspace{-2mm}}
\def\figcapdown{\vspace{-3mm}}

\def\tblcapdown{\vspace{-3mm}}

\def\tblup{\vspace{2mm}}

\def\vgap{\vspace{0mm}}

\def\G{G^*}

\def\GC{G_c}

\def\d{dist}

\def\v{v^*}
\def\GF{G_f}
\def\GB{G_b}
\def\FF{F_f}
\def\FB{F_b}
\def\HF{H_f}

\def\QF{Q_f}

\def\kf{\kappa_f}

\def\QC{Q_c}
\def\P{P^*}
\def\PA{P^\circ}

\begin{document}
\begin{sloppy}

\title{Efficient Single-Source Shortest Path and Distance Queries on Large Graphs}

\numberofauthors{1}

\author{
\alignauthor
Andy Diwen Zhu $\qquad$ Xiaokui Xiao $\qquad$ Sibo Wang $\qquad$ Wenqing Lin \\
\affaddr{
School of Computer Engineering \\
Nanyang Technological University\\
Singapore
}\\
\email{
\{dwzhu, xkxiao, swang, wlin\}@ntu.edu.sg
}
}

\maketitle
\begin{abstract}
%

This paper investigates two types of graph queries: {\em single source distance (SSD)} queries and {\em single source shortest path (SSSP)} queries. Given a node $v$ in a graph $G$, an SSD query from $v$ asks for the distance from $v$ to any other node in $G$, while an SSSP query retrieves the shortest path from $v$ to any other node. These two types of queries are fundamental building blocks of numerous graph algorithms, and they find important applications in graph analysis, especially in the computation of graph measures. Most of the existing solutions for SSD and SSSP queries, however, require that the input graph fits in the main memory, which renders them inapplicable for the massive disk-resident graphs commonly used in web and social applications. The only exceptions are a few techniques that are designed to be I/O efficient, but they all focus on undirected and/or unweighted graphs, and they only offer sub-optimal query efficiency.

To address the deficiency of existing work, this paper presents {\em Highways-on-Disk (HoD)}, a disk-based index that supports both SSD and SSSP queries on directed and weighted graphs. The key idea of HoD is to augment the input graph with a set of auxiliary edges, and exploit them during query processing to reduce I/O and computation costs. We experimentally evaluate HoD on both directed and undirected real-world graphs with up to billions of nodes and edges, and we demonstrate that HoD significantly outperforms alternative solutions in terms of query efficiency.

\end{abstract}


\section{Introduction} \label{sec:intro}


Given a graph $G$, a {\em single source distance (SSD)} query from a node $v \in G$ asks for the distance from $v$ to any other node in $G$. Meanwhile, a {\em single source shortest path (SSSP)} query retrieves the shortest path from $v$ to any other node. These two types of queries find important applications in graph analysis \cite{ckcc12}, especially in the computation of graph measures \cite{acim99,bkmm07,ew04,tk11}. For example, the estimation of {\em closeness} measures \cite{ew04} on a graph $G$ requires performing SSD queries from a large number of nodes in $G$, while the approximation of {\em betweenness} measures \cite{bkmm07} requires executing numerous SSSP queries.


\vgap


The classic solution for SSD and SSSP queries is Dijkstra's algorithm \cite{d59}. Given a SSD or SSSP query from a node $s$, Dijkstra's algorithm traverses the graph starting from $s$, such that the nodes in $G$ are visited in ascending order of their distances from $s$. Once a node $v$ is visited, the algorithm returns the distance from $s$ to $v$ based on the information maintained during the traversal; the shortest path from $s$ to $v$ can also be efficiently derived if required.

\vgap

A plethora of techniques have been proposed to improve over Dijkstra's algorithm for higher query efficiency (see \cite{dss09,s12} for surveys). Although those techniques all require pre-processing the given graph (which incurs extra overhead compared with Dijkstra's algorithm), the pre-computation pays off when the number of queries to be processed is large, as is often the case in graph analysis. Nevertheless, most of the existing techniques assume that the given graph fits in the main memory (for pre-computation and/or query processing), which renders them inapplicable for the massive disk-resident graphs commonly used in web and social applications. There are a few methods \cite{mo09,mz03,mz12,m09,ks96} that address this issue by incorporating Dijkstra's algorithm with I/O-efficient data structures, but the performance of those methods are shown to be insufficient for practical applications \cite{ckcc12}. The main reason is that, when Dijkstra's algorithm traverses the graph, the order in which it visits nodes can be drastically different from the order in which the nodes are arranged on the disk. This leads to a significant number of random disk accesses, which results in poor query performance.

\vgap

In contrast to the aforementioned techniques, Cheng et al.\ \cite{ckcc12} propose the first practically efficient index (named {\em VC-Index}) for SSD and SSSP queries on disk-resident graphs. The basic idea of VC-Index is to pre-compute a number of {\em reduced} versions of the input graph $G$. Each reduced graph contains some relatively important nodes in $G$, as well as the distances between some pairs of those nodes. During query processing, VC-Index scans a selected subset of reduced graphs, and then derives query results based on the pre-computed distances. Compared with those methods based on Dijkstra's algorithm \cite{mo09,mz03,mz12,m09,ks96}, VC-Index is more efficient as it only performs sequential reads on disk-resident data.

\vgap

\header
{\bf Motivation and Contribution.} All existing disk-based solutions for SSD and SSSP queries \cite{mo09,mz03,mz12,m09,ks96} require that the input graph is undirected, which renders them inapplicable for any application built upon directed graphs. This is rather restrictive as numerous important types of graphs (e.g., road networks, web graphs, social graphs) are directed in nature. Furthermore, even when the input graph is undirected, the query efficiency of the existing solutions is less than satisfactory. In particular, our experiments (in Section~\ref{sec:exp}) show that VC-Index, albeit being the state of the art, requires tens of seconds to answer a single SSD query on a graph with less than $100$ million edges, and needs more than two days to estimate the closeness measures on the same graph.

\vgap

To address the deficiency of existing work, this paper proposes {\em Highways-on-Disk (HoD)}, a disk-based index that supports both SSD and SSSP queries on directed and weighted graphs. The key idea of HoD is to augment the input graph with a set of auxiliary edges (referred to as {\em shortcuts} \cite{ss05}), and exploit them during query processing to reduce I/O and computation costs. For example, Figure~\ref{fig:over-steps}a illustrates a graph $G$, and Figure~\ref{fig:over-steps}e shows an augmented graph $\G$ constructed from $G$. $\G$ contains three shortcuts: $\langle v_8, v_9\rangle$, $\langle v_9, v_7\rangle$, and $\langle v_9, v_{10}\rangle$. Each shortcut has the same length with the shortest path connecting the endpoints of the shortcut. For example, the length of $\langle v_8, v_9\rangle$ equals $2$, which is identical to the length of the shortest path from $v_8$ to $v_9$. Intuitively, the shortcuts in $\G$ enable HoD to efficiently traverse from one node to another (in a manner similar to how highways facilitate traversal between distant locations). For instance, if we are to traverse from $v_1$ to $v_{10}$ in $\G$, we may follow the path $\langle v_1, v_9, v_{10}\rangle$, which consists of only three nodes; in contrast, a traversal from $v_1$ to $v_{10}$ in $G$ would require visiting five nodes: $v_1$, $v_9$, $v_6$, $v_7$, and $v_{10}$.

\vgap

In general, when HoD answers an SSD or SSSP query, it often traverses the augmented graph via shortcuts (instead of the original edges in $G$). We show that, with proper shortcut construction and index organization, the query algorithm of HoD always traverses nodes in the same order as they are arranged in the index file. Consequently, HoD can answer any SSD or SSSP query with a linear scan of the index file, and its CPU cost is linear to the number of edges in the augmented graph. 
We experimentally evaluate HoD on a variety of real-world graphs with up to $100$ million nodes and $3$ billion edges, and we demonstrate that HoD significantly outperforms VC-Index in terms of query efficiency. In particular, the query time of HoD is smaller than that of VC-Index by up to two orders of magnitude. Furthermore, HoD requires a smaller space and pre-computation time than VC-Index in most cases.

\vgap

\section{Problem Definition}\label{sec:def}


Let $G$ be a weighted and directed graph with a set $V$ of nodes and a set $E$ of edges. Each edge $e$ in $E$ is associated with a positive weight $l(e)$, which is referred to as the {\em length} of $e$. A path $P$ in $G$ is a sequence of nodes $\langle v_1, v_2, \ldots, v_k \rangle$, such that $\langle v_i, v_{i+1}\rangle$ ($i \in [1, k-1]$) is a directed edge in $G$. The length of $P$ is defined as the sum of the length of each edge on $P$. We use $l(e)$ and $l(P)$ to denote the length of an edge $e$ and a path $P$, respectively.


\vgap


For any two nodes $s$ and $t$ in $G$, we define the {\em distance} from $s$ to $t$, denoted as $\d(s, t)$, as the length of the shortest path from $s$ to $t$. Given a {\em source node} $s$ in $G$, a {\em single-source distance (SSD)} query asks for the distance from $s$ to any other node in $G$. Meanwhile, a {\em single-source shortest path (SSSP)} query from $s$ retrieves not only the distance from $s$ to any other node $v$, but also the {\em predecessor} of $v$, i.e., the node that immediately precedes $v$ in the shortest path from $s$ to $v$. Note that, given the predecessor of each node, we can easily reconstruct the shortest path from $s$ to any node $v$ by backtracking from $v$ following the predecessors. One may also consider an alternative formulation of SSD (resp.\ SSSP) query that, given only a {\em destination node} $t$, asks for the distance (resp.\ shortest path) from any other node to $t$. For simplicity, we will focus on SSD and SSSP queries from a source node $s$, but our solution can be easily extended to handle queries under the alternative formulation.


\vgap

Let $M$ be the size of the main memory available, and $B$ be the size of a disk block, both measured in the number of words. We assume that $B \le |V| \le M \le |E|$, i.e., the main memory can accommodate all nodes but not all edges in $G$. This is a realistic assumption since modern machines (even the commodity ones) have gigabytes of main memory, which is sufficient to store the node set of a graph with up to a few billion nodes. On the other hand, the number of edges in a real graph is often over an order of magnitude larger than the number of nodes, due to which $E$ can be enormous in size and does not fit in the main memory.

\vgap

Our objective is to devise an index structure on $G$ that answers any SSD or SSSP query with small I/O and CPU costs, such that the index requires at most $M$ main memory in pre-computation and query processing. In what follows, we will first focus on SSD queries in Sections \ref{sec:overview}-\ref{sec:sssp}, and will extend our solution for SSSP queries in Section~\ref{sec:exten}.

\section{Solution Overview}\label{sec:overview}

As mentioned in Section~\ref{sec:intro}, the main structure of HoD is a graph $\G$ that augments the input graph $G$ with shortcuts. In this section, we present the overall idea of how the shortcuts in $\G$ are constructed and how they can be utilized for query processing, so as to form a basis for the detailed discussions in subsequent sections.

\begin{figure*}[t]
\begin{small}
\centering
\begin{tabular}{ccccc}
\multicolumn{5}{c}{\hspace{-4mm} \includegraphics[height=3mm]{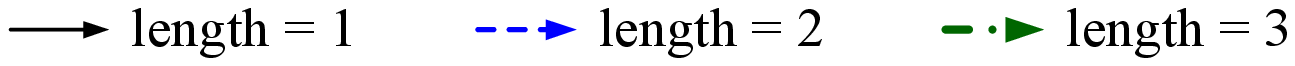}} \vspace{0mm} \\
\hspace{-2mm} \includegraphics[width=32mm]{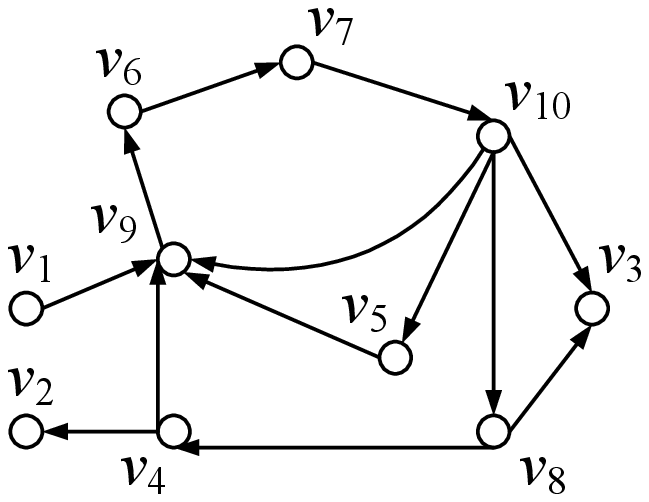}
&
\hspace{0mm} \includegraphics[width=24mm]{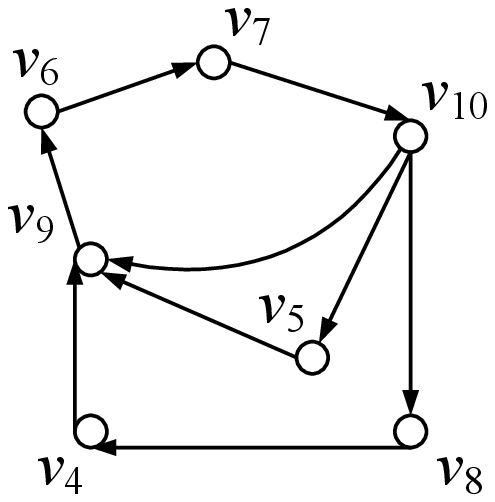}
&
\hspace{0mm} \includegraphics[width=24mm]{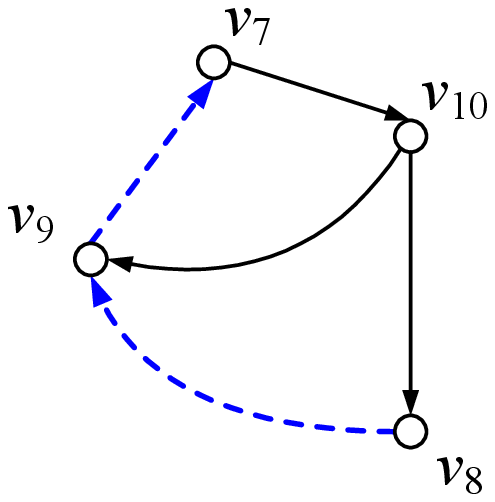}
&
\hspace{-2mm} \includegraphics[width=24mm]{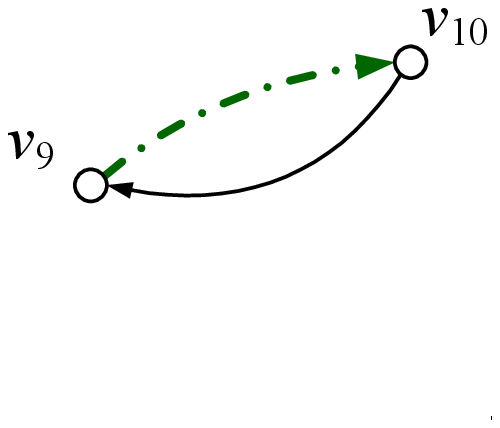}
&
\hspace{-2mm} \includegraphics[width = 47mm]{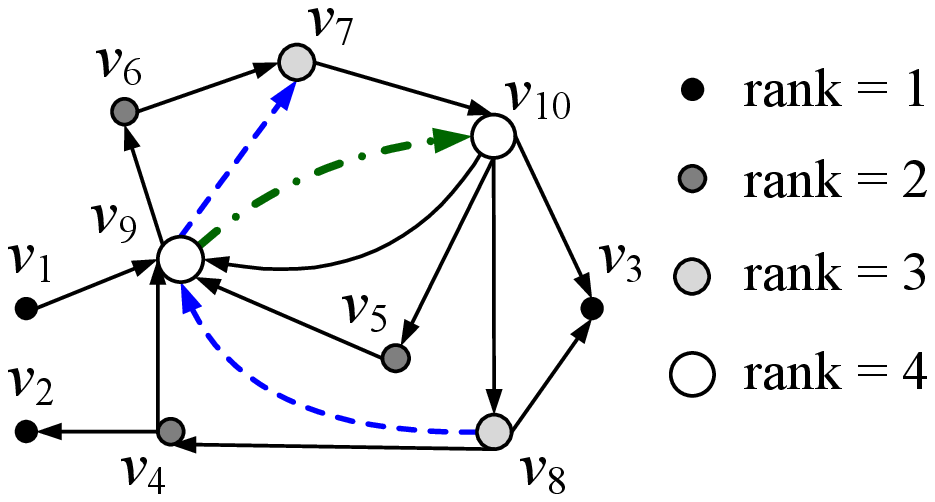}\\
\hspace{-2mm} (a) Graph $G$. & \hspace{0mm} (b) Reduced graph $G_1$. &
\hspace{0mm} (c) Reduced graph $G_2$. & \hspace{-2mm} (d) Reduced graph $G_3$ & \hspace{-5mm} (e) Augmented Graph $\G$. \\
(equivalently, $G_0$) & & & (i.e., the core graph $\GC$). &
\end{tabular}
\vspace{-4mm} \caption{Graph reduction and shortcut construction.} \figcapdown
\label{fig:over-steps}
\end{small}
\end{figure*}


\subsection{Shortcut Construction} \label{sec:overview-pre}

In a nut shell, HoD constructs shortcuts with an iterative procedure, which takes as input a copy of the graph $G$ (denoted as $G_0$). In the $i$-th ($i \ge 1$) iteration of the procedure, HoD first {\em reduces} $G_{i-1}$ by removing a selected set of {\em less important} nodes in $G_{i-1}$, and then, it constructs shortcuts in the reduced graph to ensure that the distance between any two remaining nodes is not affected by the node removal. The resulting graph (with shortcuts added) is denoted as $G_i$, and it is fed as the input of the ($i{+}1$)-th iteration of procedure. This procedure terminates only when the reduced graph $G_i$ is {\em sufficiently small}. All shortcuts created during the procedure are inserted into the original graph $G$, leading to an augmented graph $\G$ that would be used by HoD for query processing. We illustrate the iterative procedure with an example as follows.



\vgap

\begin{example} \label{example:over-steps}
\em Assume that the input to the iterative procedure is the graph $G_0$ in Figure~\ref{fig:over-steps}a. Further assume that the reduced graph is sufficiently small if it contains at most two nodes and two edges. In the first iteration of the procedure, HoD inspects $G_0$ and identifies $v_1$, $v_2$, and $v_3$ as less important nodes. To explain, observe that the node $v_1$ in $G_0$ does not have any incoming edge, while $v_2$ and $v_3$ have no outgoing edges. As a consequence, $v_1$, $v_2$, and $v_3$ do not lie on the shortest path between any two other nodes. That is, even if we remove $v_1$, $v_2$, and $v_3$ from $G_0$, the distance between any two remaining nodes is not affected. Intuitively, this indicates that $v_1$, $v_2$, $v_3$ are of little importance for SSD queries. Therefore, HoD eliminates $v_1$, $v_2$, and $v_3$ from $G_0$, which results in the reduced graph $G_1$ in Figure~\ref{fig:over-steps}b.

\vgap

In the second iteration, HoD selects $v_4$, $v_5$, and $v_6$ as the less important nodes in $G_1$, and removes them from $G_1$. The removal of $v_4$ changes the distance from $v_8$ to $v_9$ to $+\infty$, since $\langle v_8, v_4, v_9\rangle$ is the only path (in $G_1$) that starts at $v_8$ and ends at $v_9$. To mitigate this change, HoD inserts into $G_1$ a shortcut $\langle v_8, v_9\rangle$ that has the same length with $\langle v_8, v_4, v_9\rangle$, as illustrated in Figure~\ref{fig:over-steps}c. As such, the distance between any two nodes in $G_1$ remains unchanged after $v_4$ is removed. Similarly, when HoD eliminates $v_6$, it constructs a shortcut $\langle v_9, v_7\rangle$ with a length $2$ to reconnect the two neighbors of $v_6$. Meanwhile, $v_5$ is removed without creating any shortcut, since deleting $v_5$ does not change the distance between its two neighbors. Figure~\ref{fig:over-steps}c illustrates the resulting reduced graph $G_2$.

To explain why HoD chooses to remove $v_4$, $v_5$, and $v_6$ from $G_1$, observe that each of those nodes has only two neighbors. For any of such nodes, even if the removal of the node changes the distance between its neighbors, HoD only needs to construct one shortcut to reconnect its neighbors. In other words, the number of shortcuts required is minimum, which helps reduce the space consumption of HoD. In contrast, if HoD chooses to remove $v_9$ from $G_1$ (which has a larger number of neighbors than $v_4$, $v_5$, and $v_6$), then much more shortcuts would need to be constructed.

\vgap

Finally, in the third iteration, HoD removes $v_7$ and $v_8$ from $G_2$ as they are considered unimportant. The removal of $v_7$ leads to a new shortcut $\langle v_9, v_{10}\rangle$ with a length $3$, since $\langle v_9, v_7, v_{10}\rangle$ is the only path connecting $v_9$ to $v_{10}$, and the length of the path equals $3$. On the other hand, $v_8$ is directly eliminated as it is not on the shortest path between its only two neighbors $v_9$ and $v_{10}$. Figure~\ref{fig:over-steps}d shows the reduced graph $G_3$ after the removal of $v_7$ and $v_8$.

Assume that the reduced graph $G_3$ is considered sufficiently small by HoD. Then, the iterative procedure of HoD would terminate. The three shortcuts created during the procedure (i.e., $\langle v_8, v_9\rangle$, $\langle v_9, v_7 \rangle$, and $\langle v_9, v_{10}\rangle$) are added into the original graph $G$, which leads to the augmented graph $\G$ in Figure~\ref{fig:over-steps}d. \done
\end{example}

\vgap

The above discussion leaves several issues open, i.e., (i) the specific criterion for identifying less important nodes in the reduced graph, (ii) the detailed algorithm for generating shortcuts after node removal, and (iii) the exact termination condition of the reduction procedure. We will clarify these issues in Section~\ref{sec:pre} by presenting the detailed preprocessing algorithm of HoD. For the discussions in the rest of this section, it suffices to know that when HoD terminates the reduction procedure, the reduced graph must fit in the main memory. We use $\GC$ to denote this memory-resident reduced graph, and we refer to it as the {\em core graph}. (Note that $\GC$ is a subgraph of the augmented graph $\G$.) In addition, we define the {\em rank} $r(v)$ of each node $v$ in $G$ as follows:
\begin{enumerate} [itemsep = 0.5mm, topsep = 2mm] 
\item If $v$ is removed in the $i$-th iteration of the iterative procedure, then $r(v) = i$;

\item If $v$ is not removed in any iteration (i.e., $v$ is retained in the core graph $\GC$), then $r(v) = 1 + \max_{v \notin \GC} r(v)$, i.e., $r(v)$ is larger than the maximum rank of any node not in $\GC$.
\end{enumerate}
For instance, in Example~\ref{example:over-steps},
the ranks of $v_1$, $v_2$, and $v_3$ equal $1$, since they are removed from $G$ in the first iteration of the reduction procedure. Similarly, $r(v_4) = r(v_5) = r(v_6) = 2$, and $r(v_7) = r(v_8) = 3$. The ranks of $v_9$ and $v_{10}$ equal $4$, since they are in the core graph $\GC$. The ranks of the nodes are utilized in the query processing algorithms of HoD, as will be illustrated shortly. Unless otherwise specified, we use the term {\em edge} to refer to both a shortcut and an original edge in $\G$.

\subsection{Query Processing} \label{sec:overview-query}

Given an SSD query from a node $s$, HoD answers the query with two traversals of the augmented graph $\G$. The first traversal starts from $s$, and it follows only the {\em outgoing} edges of each node, ignoring any edge {\em whose starting point ranks higher than the ending point}. For instance, if HoD traverses from the node $v_9$ in Figure~\ref{fig:over-steps}e, it would ignore the outgoing edge $\langle v_9, v_7\rangle$, since $r(v_9) = 4 > r(v_7) = 3$. As such, the first traversal of HoD never moves from a high-rank node to a low-rank node, and it terminates only when no higher-rank nodes can be reached. For each node $v$ visited, HoD maintains the distance from $s$ to $v$ along the paths that have been seen during the traversal, denoted as $\d(s, v)$.

Let $V'$ be the set of nodes that are not in the core graph of $\G$. The second traversal of HoD is performed as a linear scan of the nodes in $V'$, {\em in descending order of their ranks}. For each node $v' \in V'$ scanned, HoD inspects each {\em incoming} edge $e$ of $v'$, and then checks the starting point $u$ of the edge. For any such $u$, HoD calculates $\d(s, u) + l(e)$ as an upperbound of the distance from $s$ to $v'$. (Our solution guarantees that $u$ should have been visited by HoD before $v'$.) Once all incoming edges of $v'$ are inspected, HoD derives the distance from $s$ to $v'$ based on the upperbounds, and then it moves on to the next node in $V'$. This process terminates when all nodes in $V'$ are examined.

We illustrate the above query algorithm of HoD with an example.


\vgap

\begin{example} \label{example:over-query}

\em Consider an SSD query from node $v_1$ in Figure~\ref{fig:over-steps}a. Given the augmented graph $\G$ in Figure~\ref{fig:over-steps}e, HoD first traverses $\G$ starting from $v_1$, following the outgoing edges whose ending points rank higher than the starting points. Since $v_1$ has only one outgoing edge $\langle v_1, v_9\rangle$, and since $v_9$ ranks higher than $v_1$, HoD moves from $v_1$ to $v_9$. $v_9$ has three outgoing edges: $\langle v_9, v_6\rangle$, $\langle v_9, v_7\rangle$, and $\langle v_9, v_{10}\rangle$. Among them, only $\langle v_9, v_{10}\rangle$ has an ending point that ranks higher than the starting point. Therefore, HoD moves from $v_9$ to $v_{10}$. $v_{10}$ has outgoing edges to three unvisited nodes, i.e., $v_3$, $v_5$, and $v_8$. Nevertheless, all of those nodes rank lower than $v_{10}$, and hence, they are ignored. As none of the remaining nodes can be reached without violating the constraints on node ranks, the first traversal of HoD ends. Based on the edges visited, HoD calculates $\d(v_1, v_9) = 1$ and $\d(v_1, v_{10}) = 4$.

\vgap

The second traversal of HoD examines the nodes {\em not} in the core graph in descending order of their ranks, i.e., it first examines $v_7$ and $v_8$ (whose ranks equal $3$), followed by $v_4$, $v_5$, and $v_6$ (whose ranks equal $2$), and finally $v_2$ and $v_3$ (whose ranks equal $1$), ignoring $v_1$ (as it is the source node of the query). $v_7$ has two incoming edges, $\langle v_6, v_7\rangle$ and $\langle v_9, v_7\rangle$. Among $v_6$ and $v_9$, only $v_9$ has been visited by HoD. Therefore, HoD calculates $\d(v_1, v_7) = \d(v_1, v_9) + l(\langle v_9, v_7\rangle) = 3$. Similarly, after inspecting $v_8$'s only incoming edge $\langle v_{10}, v_8\rangle$, HoD computes $\d(v_1, v_8) = \d(v_1, v_{10}) + l(\langle v_{10}, v_8\rangle) = 5$. The remaining nodes are processed in the same manner, resulting in
\vspace{-0mm}
\begin{align}
dist(v_1, v_4) \; &= \; dist(v_1, v_8) + l(\langle v_8, v_4\rangle)  \;\;\;\; = \, 6 \nonumber \\
dist(v_1, v_5) \; &= \; dist(v_1, v_{10}) + l(\langle v_{10}, v_5\rangle) \; = \, 5 \nonumber \\
dist(v_1, v_6) \; &= \; dist(v_1, v_{9}) + l(\langle v_{9}, v_6\rangle) \;\;\;\, = \; 2 \nonumber \\
dist(v_1, v_2) \; &= \; dist(v_1, v_{4}) + l(\langle v_{4}, v_1\rangle) \;\;\;\, = \; 7. \nonumber
\end{align}
\vspace{-0mm}
Observe that all the above distances computed from $\G$ are identical with those from the original graph in Figure~\ref{fig:over-steps}a. \done

\vgap

\end{example}

\vgap

The query algorithm of HoD has an interesting property: the first traversal of the algorithm always visits nodes in ascending order of their ranks (as it never follows an edge that connects a high-rank node to low-rank node), while the second phase always visits nodes in descending rank order. Intuitively, if we maintain two copies of the augmented graph, such that the first (resp.\ second) copy stores nodes in ascending (resp.\ descending) order of their ranks, then HoD can answer any SSD query with a linear scan of the two copies. This leads to high query efficiency as it avoids random disk accesses. In Section~\ref{sec:pre}, we will elaborate how such two copies of the augmented graph can be constructed.

\section{Index Construction}\label{sec:pre}

As discussed in Section~\ref{sec:overview-pre}, the preprocessing algorithm of HoD takes as input a copy $G_0$ of the graph $G$, and it iteratively reduces $G_0$ into smaller graphs $G_1, G_2, \ldots$, during which it creates shortcuts to augment $G$. More specifically, the ($i{+}1$)-th ($i \ge 0$) iteration of the algorithm has four steps:
\begin{enumerate} [itemsep = 0.5mm, topsep = 2mm] 
\item Select a set $R_i$ of nodes to be removed from $G_{i}$.

\item For each node $v \in R_i$, construct shortcuts in $G_{i}$ to ensure that the removal of $v$ does not change the distance between any two remaining nodes.

\item Remove the nodes in $R_i$ from $G_{i}$ to obtain a further reduced graph $G_{i+1}$. Store information about the removed nodes in the index file of HoD.

\item Pass the $G_{i+1}$ to the ($i{+}2$)-th iteration as input.
\end{enumerate}
In the following, we first elaborate Steps $2$ and $3$, and then clarify Step $1$. After that, we will discuss the termination condition of the preprocessing algorithm, as well as its space and time complexities.

For ease of exposition, we represent each edge $e = \langle u, v \rangle$ as a triplet $\left\langle u, v, l(e)\right\rangle$ or $\left\langle v, u, -l(e)\right\rangle$, where $l(e)$ is the length of $e$. For example, the edge $\langle v_9, v_7 \rangle$ in Figure~\ref{fig:over-steps}a can be represented as either $\left\langle v_9, v_7, 2\right\rangle$ or $\left\langle v_7, v_9, -2\right\rangle$. That is, a negative length in the triplet indicates that the second node in the triplet is the starting point of the edge. In addition, we assume that the input graph $G$ is stored on the disk as adjacency lists, such that (i) for any two nodes $v_i$ and $v_j$, the adjacency list of $v_i$ precedes that of $v_j$ if $i < j$, and (ii) each edge $\langle v_i, v_j\rangle$ with length $l$ is stored twice: once in the adjacency list of $v_i$ (as a triplet $\langle v_i, v_j, l\rangle$), and another in the adjacency list of $v_j$ (as a triplet $\langle v_j, v_i, -l\rangle$).

\subsection{Node Removal and Shortcut Generation} \label{sec:pre-removal}

Let $\v$ be a node to be removed from $G_{i}$. We define an {\em outgoing neighbor} of $\v$ as a node $u$ to which $\v$ has an outgoing edge. Similarly, an {\em incoming neighbor} of $\v$ is a node $w$ from which $\v$ has an incoming edge. We have the following observation:


\vspace{-1mm}
\begin{observation} \label{observ:pre-neighbor}
\em
For any two nodes $v_j$ and $v_k$ in $G_i$, the distance from $v_j$ to $v_k$ changes after $\v$ is removed, if and only if the shortest path from $v_j$ to $v_k$ contains a sub-path $\langle u, \v, w\rangle$, such that $u$ (resp.\ $w$) is an incoming (resp.\ outgoing) neighbor of $\v$. \done
\end{observation}
\vspace{-1mm}

By Observation~\ref{observ:pre-neighbor}, we can preserve the distance between any two nodes in $G_i$ after removing $\v$, as long as we ensure that the distance between any incoming neighbor and any outgoing neighbor of $\v$ remains unchanged. This can be achieved by connecting the incoming and outgoing neighbors of $\v$ with shortcuts, as demonstrated in Section~\ref{sec:overview-pre}. Towards this end, a straightforward approach is to generate a shortcut $\langle u, w\rangle$ for any incoming neighbor $u$ and any outgoing neighbor $w$. The shortcuts thus generated, however, are often {\em redundant}. For example, consider the graph $G_i$ in Figure~\ref{fig:pre-contract}a. Suppose that we are to remove $v_2$, which has an incoming neighbor $v_1$ and an outgoing neighbor $v_3$. If we construct a shortcut from $v_1$ to $v_3$, it is useless since (i) $v_1$ already has an outgoing edge to $v_3$, and (ii) the edge $\langle v_1, v_3\rangle$ is even shorter than the path from $v_1$ to $v_3$ via $v_2$. As another example, assume that $v_4$ in Figure~\ref{fig:pre-contract}a is also to be removed. $v_4$ has an incoming neighbor $v_1$ and an outgoing neighbor $v_5$, but the path $\langle v_1, v_4, v_5\rangle$ is no shorter than another path from $v_1$ to $v_5$, i.e., $\langle v_1, v_3, v_5\rangle$, which does not go through $v_4$. As a consequence, even if we remove $v_4$ from $G_i$, the distance from $v_1$ to $v_5$ is still retained, and hence, it is unnecessary to insert a shortcut from $v_1$ to $v_5$.


In general, for any incoming neighbor $u$ and outgoing neighbor $w$ of $\v$, a shortcut from $u$ to $w$ is unnecessary if there is a path $P$ from $u$ to $v$, such that (i) $P$ does not go through $\v$, and (ii) $P$ is no longer than $\langle u, \v, w\rangle$. To check whether such a path $P$ exists, one may apply Dijkstra's algorithm to traverse $G_i$ from $u$ (or $w$), ignoring $\v$ during the traversal. However, when $G_i$ does not fit in main memory (as is often the case in the pre-computation process of HoD), this approach incurs significant overhead, due to the inefficiency of Dijkstra's algorithm for disk-resident graphs (as discussed in Section~\ref{sec:intro}). To address this issue, we adopt a heuristic approach that is not as effective (in avoiding redundant shortcuts) but much more efficient. Specifically, for each $\v$ in the node set $R_i$ to be removed from $G_i$, we generate a {\em candidate edge} $e_c = \langle u, w\rangle$ from each incoming neighbor $u$ of $\v$ to each outgoing neighbor $w$ of $\v$, setting the length of the shortcut to $l(\langle u, \v, w\rangle)$. For any such candidate edge $e_c$, we insert it into a temporary file $T$ as two triplets: $\langle u, w, l(e_c)\rangle$ and $\langle w, u, -l(e_c)\rangle$.

In addition to the candidate edges, we also insert two additional groups of edges (referred to as {\em baseline edges}) into the temporary file $T$ as triplets. The first group consists of any edge $e$ in $G_i$ connecting two nodes not in $R_i$, i.e., the two endpoints of $e$ are not to be removed. The second group is generated as follows: for each node $v$ not in $R_i$, we select $v$'s certain incoming neighbor $u'$ and outgoing neighbor $w'$, and we construct a baseline edge $\langle u', w'\rangle$, setting its length to $l(\langle u', v, w'\rangle)$.

The purpose of inserting a baseline edge $e$ into the temporary file $T$ is to help eliminate any redundant candidate edge that (i) shares the same endpoints with $e$ but (ii) is not shorter than $e$. Towards this end, once all baseline edges are added into $T$, we sort the triplets in $T$ using a standard algorithm for external sort, such that a triplet $t_1 = \langle v_a, v_b, l_1\rangle$ precedes another triplet $t_2 = \langle v_\alpha, v_\beta, l_2\rangle$, if any of the following conditions hold:
\begin{enumerate} [itemsep = 0.5mm, topsep = 2mm]
\item $a < \alpha$, or $a = \alpha$ but $b < \beta$.

\item $a = \alpha$, $b = \beta$, and $l_1 > 0 > l_2$. That is, any outgoing edge of a node precedes its incoming edges.

\item $a = \alpha$, $b = \beta$, $l_1 \cdot l_2 > 0$ (i.e., $t_1$ and $t_2$ are both incoming edges or both outgoing edges), and $|l_1| < |l_2|$. That is, $t_1$ and $t_2$ share the same starting and ending points, but $t_1$ is shorter than $t_2$.

\item $a = \alpha$, $b = \beta$, $l_1 \cdot l_2 > 0$, $|l_1| = |l_2|$, and $t_1$ is a baseline edge while $t_2$ is a candidate edge.
\end{enumerate}
Once $T$ is sorted, the outgoing (resp.\ incoming) edges with the same endpoints are grouped together, and the first edge in each group should have the smallest length within the group. If the first edge $e$ in a group
is a candidate edge, then we retain $e$ as it is shorter than any other baseline or candidate edges that we have generated. On the other hand, if $e$ is a baseline edge, then the distance between the endpoints of $e$ must not be affected by the removal of any nodes in $R_i$. In that case, all candidate edges in the group can be omitted. With one linear scan of the sorted $T$ and the adjacency lists of $G_i$, we can remove the information of any node in $R_i$, and merge the retained candidate edges into the adjacency lists of the remaining nodes. We illustrate the above algorithm with an example.

\begin{example} \label{example:pre-removal}
\em
Suppose that, given the graph $G_i$ in Figure~\ref{fig:pre-contract}a, we are to remove a node set $R_i = \{v_2, v_4\}$ from $G_i$. $v_2$ has only one incoming neighbor $v_1$ and one outgoing neighbor $v_3$, and $l(\langle v_1, v_2, v_3\rangle) = 2$. Accordingly, HoD generates a candidate edge $\langle v_1, v_3\rangle$ by inserting into the temporary file $T$ two triplets, $\langle v_1, v_3, 2\rangle$ and $\langle v_3, v_1, -2\rangle$. Similarly, for $v_4$, HoD creates a candidate edge $\langle v_1, v_5\rangle$, which is represented as two triplets in $T$: $\langle v_1, v_5, 2\rangle$ and $\langle v_5, v_1, -2\rangle$.

Meanwhile, the edge $\langle v_1, v_3\rangle$ in $G_i$ is selected as a baseline edge and is inserted into $T$, since neither $v_1$ nor $v_3$ is in $R_i$. In addition, HoD also generates a baseline edge $\langle v_1, v_5\rangle$ from the neighbors of $v_3$. This is because that (i) $v_1, v_3, v_5$ are not in $R_i$, (ii) $v_1$ is an incoming neighbor of $v_3$, and (iii) $v_5$ is an outgoing neighbor of $v_3$. Figure~\ref{fig:pre-contract}c illustrates the temporary file $T$ after all candidate and baseline edges are inserted, with some triplets omitted for simplicity. Figure~\ref{fig:pre-contract}d shows the file $T$ after it is sorted. The baseline edge $\langle v_1, v_3, 1\rangle$ precedes the candidate edge $\langle v_1, v_3, 2\rangle$, which indicates that we do not need to add a shortcut from $v_1$ to $v_3$. Similarly, the baseline edge $\langle v_3, v_1, -1\rangle$ precedes the candidate edge $\langle v_3, v_1, -2\rangle$, in which case the latter is omitted. Overall, each of the candidate edges in $T$ is preceded by a baseline edge, and hence, no shortcut will be created. Consequently, HoD removes from $G_i$ the adjacency lists of $v_2$ and $v_4$, as well as all edges to and from $v_2$ and $v_4$ in any other adjacency lists. This results in the reduced graph illustrated in Figure~\ref{fig:pre-contract}b. \done
\end{example}

In summary, HoD decides whether a candidate edge $e$ should be retained, by comparing it with all edges in $G_i$ as well as some two-hop paths in $G_i$. This heuristic approach may retain unnecessary candidate edges, but it does not affect the correctness of SSD queries. To understand this, recall that each candidate edge $e = \langle u, w\rangle$ has the same length with a certain path $\langle u, \v, w\rangle$ that {\em exists} in $G_i$, where $\v$ is the node whose removal leads to the creation of $e$. In other words, the length of $e$ is at least larger than or equal to the distance from $u$ to $w$. Adding such an edge into $G_i$ would not decrease the distance between any two nodes in $G_i$, and hence, retaining $e$ does not change the results of any SSD queries.

The above discussions assume that HoD has selected a set $R_i$ of nodes to be removed from $G_i$, and has decided which baseline edges are to be generated from the neighbors of the nodes not in $R_i$. We will clarify these two issues in Section \ref{sec:pre-select} and \ref{sec:pre-baseline}.


\subsection{Selecting Nodes for Removal} \label{sec:pre-select}

Consider any node $v$ in $G_i$. Intuitively, if the removal of $v$ requires us to insert a large number of shortcuts into $G_i$, then $v$ may lie on the shortest paths between many pair of nodes, in which case $v$ should be considered important. Let $B_{in}$ and $B_{out}$ be the set of incoming and outgoing neighbors of $v$, respectively. The maximum number of shortcuts induced by $v$'s removal is:
\begin{equation} \label{eqn:pre-score}
s(v) = {\big |}B_{in}{\big |} \cdot {\big |}B_{out} \setminus B_{in}{\big |} + {\big |}B_{out}{\big |} \cdot {\big |}B_{in} \setminus B_{out}{\big |}.
\end{equation}
We refer to $s(v)$ as the {\em score} of $v$ in $G_i$, and we consider $v$ {\em unimportant} if $s(v)$ is no more than the median score in $G_i$. (For practical efficiency, we use an approximated value of the median score computed from a sample set of the nodes.)


Ideally, we would like to remove all unimportant nodes from $G_i$, but this is not always feasible. To explain, consider that we are given the reduced graph $G_1$ in Figure~\ref{fig:over-steps}b, and we aim to eliminate both $v_6$ and $v_7$. $v_6$ has only one incoming neighbor $v_9$ and one outgoing neighbor $v_7$, and hence, HoD creates one candidate edge $\langle v_9, v_7\rangle$, setting its length to $2$ (i.e., the length of the path $\langle v_9, v_6, v_7\rangle$). Similarly, for $v_7$, HoD generates a candidate edge $\langle v_6, v_{10}\rangle$. These two candidate edges are intended to preserve the distance between any two nodes in $G_i$ after $v_6$ and $v_7$ are removed. However, none of the two candidate edges is valid if both $v_6$ and $v_7$ are eliminated. In particular, $\langle v_9, v_7\rangle$ points from $v_9$ to $v_7$, i.e., it connects $v_9$ to a node that no longer exists. To avoid the above error, whenever HoD chooses to delete a node $v$ from $G_i$, it will retain all neighbors of $v$ in $G_i$, even if some neighbor might be unimportant. For example, in Figure~\ref{fig:over-steps}b, if HoD decides to remove $v_6$, then it will prevent $v_7$ from being removed at the same time, and vice versa.


\begin{figure}[t]
\begin{small}
\centering
\begin{tabular}{ccc}
\multicolumn{3}{c}{\includegraphics[height=3mm]{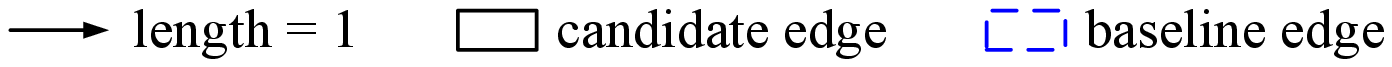}} \vspace{1mm}\\
\includegraphics[width = 15mm]{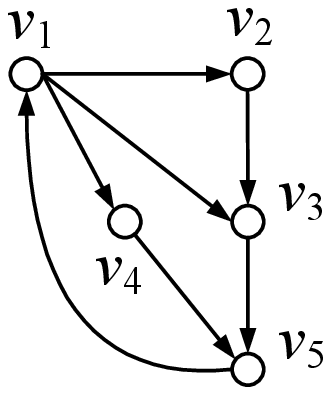} & \multirow{3}{*}[16.5mm]{\includegraphics[width = 14mm]{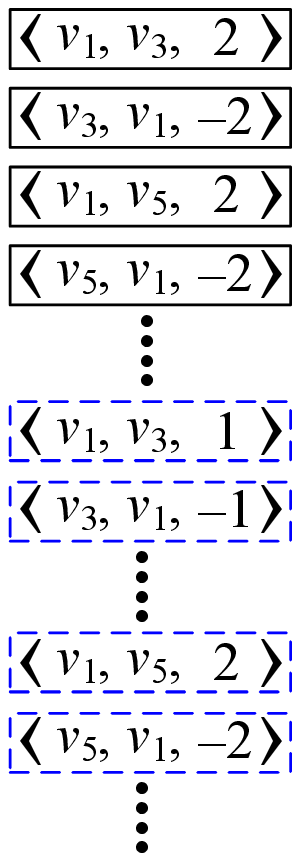}} & \multirow{3}{*}[16.5mm]{\includegraphics[width = 14mm]{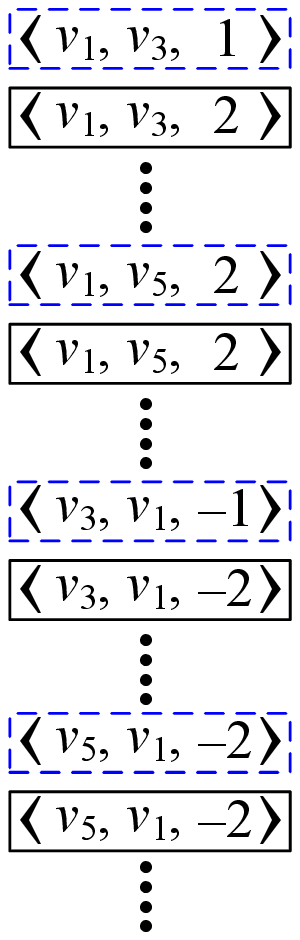}} \vspace{1mm}\\
(a) Graph $G_i$. & & \vspace{4mm}\\
\includegraphics[width = 14mm]{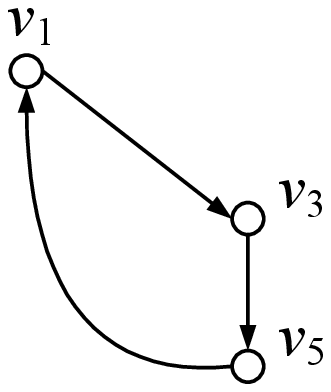} & & \\
(b) $v_2, v_4$ removed. & \hspace{3mm} (c) Before sorting. & \hspace{3mm} (d) After sorting. \\
\end{tabular}
\figcapup \caption{Node removal and shortcut generation.} \figcapdown \vspace{-1mm}
\label{fig:pre-contract}
\end{small}
\end{figure}

\subsection{Generation of Baseline Edges} \label{sec:pre-baseline}

As mentioned in Section~\ref{sec:pre-removal}, a baseline edge generated by the preprocessing algorithm of HoD is either (i) an edge in $G_i$ whose endpoints are not to be removed, or (ii) an artificial edge $\langle u, w\rangle$ that corresponds to certain two-hop path $\langle u, v, w\rangle$ in $G_i$, such that none of $u, v, w$ is to be removed. The construction of baseline edges from two-hop paths is worth discussing. First, given that there exists an enormous number of two-hop paths in $G_i$, it is prohibitive to convert each two-hop path into a baseline edge. Therefore, we only select a subset of the two-hop paths in $G_i$ for baseline edge generation. In particular, the total number of two-hop paths selected is set to $c \cdot \sum_{v \in R_i} s(v)$, where $c$ is a small constant, $s(v)$ is as defined in Equation~\ref{eqn:pre-score}, and $\sum_{v \in R_i} s(v)$ is the total number of candidate edges induced by the removal of nodes in $R_i$. In other words, we require that the number of baseline edges generated from two-hop paths is at most $c$ times the number of candidate edges. In our implementation of HoD, we set $c = 5$.


Those $c \cdot \sum_{v \in R_i} s(v)$ baseline edges are generated as follows. First, we randomly choose an edge in $G_i$, and arbitrarily select an endpoint $v$ of the edge that is not in $R_i$. (Note that such an endpoint always exists.) After that, from the incoming (resp.\ outgoing) neighbors of $v$, we randomly select a node $u$ (resp.\ $w$), and we generate a baseline edge $\langle u, w\rangle$, setting its length to $l(\langle u, v, w\rangle)$. As such, if a node $v$ is adjacent to a large number of edges, then it has a high chance of being selected to produce baseline edges. This is intuitive since such a node $v$ tends to lie on the shortest paths between many pairs of nodes, and hence, the baseline edges generated from $v$ may be more effective in eliminating redundant shortcuts.

\subsection{Termination Condition} \label{sec:pre-terminate}

As mentioned in Section~\ref{sec:overview}, HoD requires that the core graph $\GC$ fits in the main memory, where $\GC$ is the reduced graph obtained in the last iteration of HoD's preprocessing algorithm. Accordingly, we do not allow the pre-computation procedure of HoD to terminate before the reduced graph $G_i$ has a size no more than $M$. In addition, even after $G_i$ fits in the main memory, we will still continue the preprocessing procedure, until the size of $G_i$ is reduced by less than $5\%$ in an iteration of the processing algorithm. This is intended to reduce the size of the core graph $\GC$ to improve query efficiency, as will be explained in Section~\ref{sec:sssp}.

\begin{figure}[t]
\centering
\begin{small}
\begin{tabular}{ccc}
\hspace{-3mm}
\includegraphics[width = 30mm]{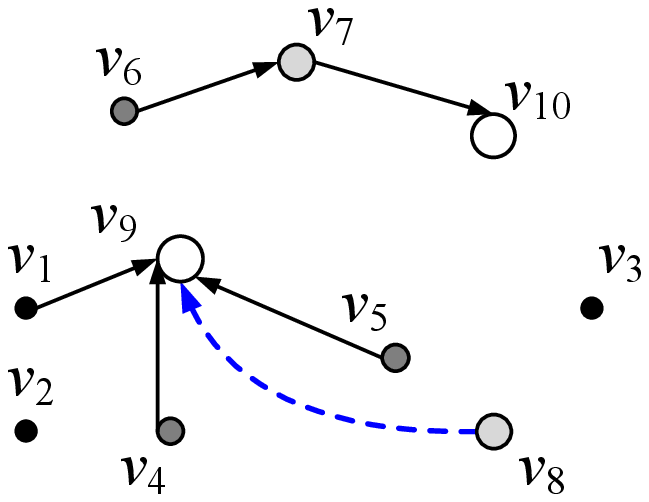} &
\hspace{-3mm}
\includegraphics[width = 18mm]{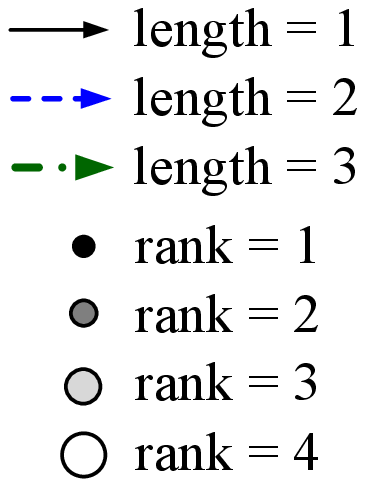} &
\includegraphics[width = 30mm]{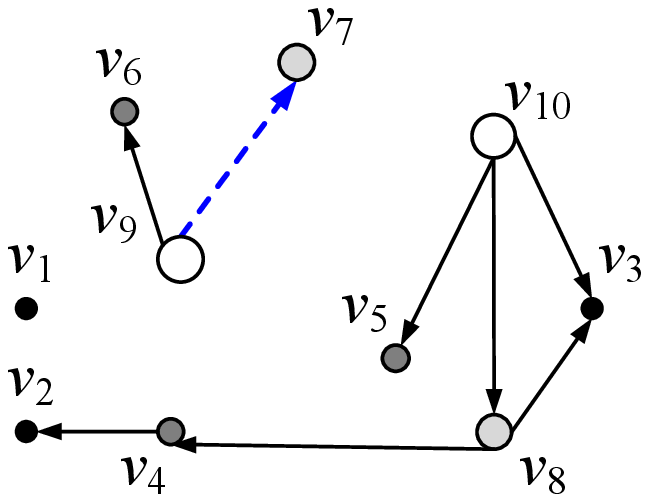} \\
\hspace{-3mm} (a) $\GF$. & \multicolumn{2}{c}{\hspace{15mm} (b) $\GB$.}
\end{tabular}
\figcapup \vspace{-2mm} \caption{Forward graph $\GF$ and backward graph $\GB$.} \figcapdown \vspace{-1mm}
\label{fig:pre-graphs}
\end{small}
\end{figure}

\subsection{Index File Organization} \label{sec:pre-file}


Once the preprocessing procedure completes, the core graph $\GC$ is written to the disk in the form of adjacency lists. Meanwhile, the adjacency list of each node not in $\GC$ is separated into two parts that are stored in two different files, $\FF$ and $\FB$. These two files are created at the beginning of HoD's preprocessing algorithm, and they are initially empty. Whenever a node $v$ is removed from the reduced graph $G_i$, we inspect the adjacency list of $v$ in $G_i$, and we append all of $v$'s outgoing (resp.\ incoming) edges to $\FF$ (resp.\ $\FB$). Upon termination of the preprocessing procedure, we reverse the order of nodes in $\FB$, but retain the order of nodes in $\FF$. We refer to the graph represented by $\FF$ as the {\em forward graph}, denoted as $\GF$. Meanwhile, we refer to the graph represented by $\FB$ as the {\em backward graph}, denoted as $\GB$. When combined, $\GC$, $\GF$, and $\GB$ form the augmented graph that is used by HoD for query processing. For example, for the augmented graph in Figure~\ref{fig:over-steps}a, its core graph is as illustrated in Figure~\ref{fig:over-steps}e, while its forward and backward graphs are as shown in Figure~\ref{fig:pre-graphs}. For ease of exposition, we will abuse notation and use $\GF$ (resp.\ $\GB$) to refer to both $\GF$ (resp.\ $\GB$) and its underlying file structure $\FF$ (resp.\ $\FB$).



$\GF$ and $\GB$ have two interesting properties. First, all nodes in $\GF$ (resp.\ $\GB$) are stored in ascending (resp.\ descending) order of their ranks. To explain this, recall that any node $v$ removed in the $i$-th iteration of the preprocessing algorithm has a rank $r(v)=i$. Consequently, if a node $u$ is stored in $\GF$ before another node $w$, then $r(u) \le r(w)$. As for $\GB$, since we reverse the order of all edges in $\GB$ upon termination of the preprocessing produce, we have $r(u) \ge r(w)$ for any node $u$ that precedes another node $w$ in $\GB$. Second, for any node $v$, its edges in $\GF$ and $\GB$ only connect it to the nodes whose ranks are strictly higher than $v$. This is because, by the time $v$ is removed from the reduced graph, all nodes that rank lower than $v$ must have been eliminated from the reduced graph, and hence, any edge in $v$'s adjacency list only links $v$ to the nodes whose rank is at least $r(v)$. Meanwhile, any neighbor $u$ of $v$ in the reduced graph should have a rank higher than $r(v)$. Otherwise, we have $r(u) = r(v)$, which, by the definition of node ranks, indicates that $u$ and $v$ are removed in the same iteration of the preprocessing algorithm. This is impossible as the pre-computation procedure of HoD never eliminates two adjacent nodes in the same iteration, as explained in Section~\ref{sec:pre-removal}. In Section~\ref{sec:sssp}, we will show how HoD exploits the above two properties of $\GF$ and $\GB$ to efficiently process SSD queries.



\subsection{Cost Analysis} \label{sec:pre-complexity}

The preprocessing algorithm of HoD requires $O(n)$ main memory, where $n$ is the number of nodes in $G$. This is because (i) when removing a node $v$ from the reduced graph, HoD needs to record the neighbors of $v$ and exclude them from the node removal process, and (ii) $v$ may have $O(n)$ neighbors. Other parts of the preprocessing algorithm do not have a significant memory requirement.

The major I/O and CPU costs of the preprocessing algorithm are incurred by sorting the edge triplets in each iteration. In the worst case when the input graph $G$ is a complete graph, there are $O(n^2)$ triplets generated in each iteration, leading to a prohibitive I/O and CPU overhead. Fortunately, real-world graphs are seldom complete graphs, and they tend to contain a large number of nodes with small degrees. In that case, each iteration of HoD's preprocessing procedure would only generates a moderate number of edge triplets, leading to a relatively small overhead.

Lastly, the space consumption of HoD's index is $O(n^2)$ since, in the worst case, HoD may construct a shortcut from each node $v$ to every node that ranks higher than $v$. This space complexity is unfavorable, but it is comparable to the space complexity of VC-Index \cite{ckcc12}. In addition, as shown in our experiments, the space requirement of HoD in practice is significantly smaller than the worst-case bound.

\section{Algorithm for SSD Queries} \label{sec:sssp}

Given an SSD query from a node $s$ that is not in the core graph $\GC$, HoD processes the query in three steps:
\begin{enumerate}[itemsep = 0.5mm, topsep = 2mm]
\item {\bf Forward Search:} HoD traverses the forward graph $\GF$ starting from $s$, and for each node $v$ visited, computes the distance from $s$ to $v$ in $\GF$.

\item {\bf Core Search:} HoD reads the core graph $\GC$ into the main memory, and continues the forward search by following the outgoing edges of each node in $\GC$.

\item {\bf Backward Search:} HoD linearly scans the backward graph $\GB$ to derive the exact distance from $s$ to any node not in $\GC$.
\end{enumerate}
On the other hand, if $s$ is in $\GC$, then HoD would answer the query with a core search followed by a backward search, skipping the forward search. In what follows, we will present the details of three searches performed by HoD. For convenience, we define an {\em index} value $\theta(v)$ for each node {\em not} in the core graph $\GC$, such that $\theta(v) = i$ only if the $i$-th adjacency list in $\GF$ belongs to $v$. By the way $\GF$ is constructed (see Section~\ref{sec:pre-file}), for any two nodes $u$ and $v$ with $\theta(u) < \theta(v)$, the rank of $u$ is no larger than the rank of $v$.

\subsection{Forward Search} \label{sec:sssp-forward}

The forward search of HoD maintains a hash table $\HF$ and a min-heap $\QF$. In particular, $\HF$ maps each node $v$ to a key $\kf(v)$, which equals the length of the shortest path from $s$ to $v$ that is seen so far. Initially, $\kf(s) = 0$, and $\kf(v) = +\infty$ for any node $v \ne s$. On the other hand, each entry in $\QF$ corresponds to a node $v$, and the key of the entry equals $\theta(v)$, i.e., the index of $v$. As will become evident shortly, $\QF$ ensures that the forward search visits nodes in ascending order of their indices, and hence, it scans the file structure of $\GF$ only once, without the need to backtrack to any disk block that it has visited before.

HoD starts the forward search by inspecting each edge $e = \langle s, v\rangle$ adjacent to $s$ in $\GF$, and then inserting $v$ into $\HF$ with a key $\kappa(v) = l(e)$. (Note that $\GF$ contains only outgoing edges.) In addition, HoD also inserts $v$ into $\QF$. After that, HoD iteratively removes the top entry $u$ in $\QF$, and processes $u$ as follows: for each edge $e = \langle u, v\rangle$ adjacent to $u$, if $\kf(v) = +\infty$ in the hash table $\HF$, HoD sets $\kf(v) = \kf(u) + l(e)$ and inserts $v$ into $\QF$; otherwise, HoD sets $\kf(v) = \min\{\kf(v), \kf(u) + l(e)\}$. When $\QF$ becomes empty, HoD terminates the forward search, and retains the hash table $\HF$ for the second step of the algorithm, i.e., the core search.

\subsection{Core Search} \label{sec:sssp-core}

The core search of HoD is a continuation of the forward search, and it inherits the hash table $\HF$ created during the forward search. In addition to $\HF$, HoD creates a min-heap $\QC$, such that $\QC$ stores entries of the form $\langle v, \kf(v)\rangle$, where $v$ is a node and $\kf(v)$ is the key of $v$ in $\HF$. For any node $u$ with $\kf(u) \ne +\infty$ (i.e., $u$ is visited by the forward search), HoD inserts $u$ into $\QC$.

Given $\HF$ and $\QC$, HoD performs the core search in iterations. In each iteration, HoD extracts the top entry $v$ from $\QC$, and examines each outgoing edge $e$ of $v$. For every such edge, HoD inspects its ending point $w$, and sets $\kf(w) = \min\{\kf(w), \kf(v) + l(e)\}$. Then, HoD adds $w$ into $\QC$ if $w$ is currently not in $\QC$. This iterative procedure is repeated until $\QC$ becomes empty. After that, the hash table $\HF$ is sent to the last step (i.e., the backward search) for further processing.

\subsection{Backward Search} \label{sec:sssp-backward}

Given the hash table $\HF$ obtained from the core search, the reversed search of HoD is performed as a sequential scan of the backward graph $\GB$, which stores nodes in descending order of their index values. For each node $v$ visited during the sequential scan, HoD checks each edge $e = \langle u, v\rangle$ adjacent to $v$. (Note that $\GB$ contains only incoming edges). If $\kf(u) \ne +\infty$ and $\kf(u) + l(e) < \kf(v)$, then HoD sets $\kf(v) = \kf(u) + l(e)$. Once all nodes in $\GB$ are scanned, HoD terminates the backward search and, for each node $v$, returns $\kf(v)$ as the distance from $s$ to $v$.

One interesting fact about the backward search is that it does not require a heap to decide the order in which the nodes are visited. This leads to a much smaller CPU cost compared with Dijkstra's algorithm, as it avoids all expensive heap operations.

\subsection{Correctness and Complexities} \label{sec:sssp-proof}

Compared with Dijkstra's algorithm, the main difference of HoD's query algorithm is that it visits nodes in a pre-defined order based on their ranks. The correctness of this approach is ensured by the shortcuts constructed by the preprocessing algorithm of HoD. In particular, for any two nodes $s$ and $t$ in $G$, it can be proved that the augmented graph $\G$ always contains a path $P$ from $s$ to $t$, such that (i) $P$'s length equals the distance from $s$ to $t$ in $G$, and (ii) $P$ can be identified by HoD with a forward search from $s$, followed by a core search and a backward search. More formally, we have the following theorem.
\begin{theorem} \label{thrm:sssp-correct}
Given a source node $s$, the SSD query algorithm of HoD returns $\d(s, v)$ for each node $v \in G$.
\end{theorem}
The proof of Theorem~\ref{thrm:sssp-correct} in included in Appendix~\ref{sec:proofs}.

The query algorithm of HoD requires $O(n + m_c)$ main memory, where $m_c$ is the size of the core graph. This is due to the fact that (i) the forward, core, and back searches of HoD all maintain a hash table that takes $O(n)$ space, and (ii) the core search requires reading the core graph $\GC$ into the memory. The time complexity of the algorithm is $O(n \log n + m')$, where $m'$ is the total number of edges in $\GC$, $\GF$, and $\GB$. The reason is that, when processing an SSD query, HoD may need to scan $\GF$, $\GC$, and $\GB$ once, and it may need to put $O(n)$ nodes into a min-heap. Finally, the I/O costs of the algorithm is $O((n + m')/B)$, since it requires at most one scan of $\GF$, $\GC$, and $\GB$.

\section{Extension for SSSP Queries} \label{sec:exten}

Given a source node $s$, an SSSP query differs from an SSD query only in that it asks for not only (i) the distance from $s$ to any other node $v$, but also (ii) the predecessor of $v$, i.e., the node that immediately precedes $v$ on the shortest path from $s$ to $v$. To extend HoD for SSSP queries, we associate each edge $\langle u, w\rangle$ in the augment graph $\G$ with a node $v$, such that $v$ immediately precedes $w$ on the shortest path from $u$ to $w$ in $G$. For example, given the augmented graph in Figure~\ref{fig:over-steps}e, we would associate the edge $\langle v_9, v_7\rangle$ with $v_6$, since (i) the shortest path from $v_9$ to $v_7$ in $G$ is $\langle v_9, v_6, v_7\rangle$, and (ii) $v_6$ immediately precedes $v_7$ in the path.

With the above extension, HoD processes any SSSP query from a node $s$ using the algorithm for SSD query with one modification: Whenever HoD traverses an edge $\langle u, w\rangle$ and finds that $\d(s, u) + l(\langle u, w \rangle) < \d(s, w)$, HoD would not only update $\d(s, w)$ but also record the node associated with $\langle u, w\rangle$. That is, for each node $w$ visited, HoD keeps track of the predecessor of $w$ in the shortest path from $s$ to $w$ that have been seen so far. As such, when the SSD query algorithm terminates, HoD can immediately return $\d(s, w)$ as well as the predecessor of $w$.

Finally, we clarify how the preprocessing algorithm of HoD can be extended to derive the node associated with each edge. First, for each edge $e$ in the original graph, HoD associates $e$ with the starting point of $e$. After that, whenever HoD generates a candidate edge $\langle u, w\rangle$ during the removal of a node $v$, HoD would associate $\langle u, w\rangle$ with the node that is associated with the edge $\langle v, w\rangle$. For example, in Figure~\ref{fig:over-steps}c, when HoD removes $v_7$ and creates a candidate edge $\langle v_9, v_{10}\rangle$, it associates the edge with $v_7$, which is the node associated with $\langle v_7, v_{10}\rangle$.

\section{Experiments}\label{sec:exp}


This section experimentally compares HoD with three methods: (i) VC-Index \cite{ckcc12}, the state-of-the-art approach for SSD and SSSP queries on disk-resident graphs; (ii) EM-BFS \cite{adm06}, an I/O efficient method for breadth first search; and (iii) EM-Dijk \cite{mo09}, an I/O efficient version of Dijkstra's algorithm. We include EM-BFS since, on unweighted graphs, any SSD query can be answered using breadth first search, which is generally more efficient than Dijkstra's algorithm. We obtain the C++ implementations of VC-Index, EM-BFS, and EM-Dijk from their inventors, and we implement HoD with C++. As the implementation of VC-Index only supports SSD queries, we will focus on SSD queries instead of SSSP queries. All of our experiments are conducted on a machine with a $2.4$GHz CPU and $32$GB memory.

\begin{table}[t]
\vspace{-2mm}
\centering
\begin{small}
\caption{Datasets.}
\label{tbl:exp-data}
\tblup
\begin{tabular}{|c|c|c|@{\hspace{1mm}}c@{\hspace{1mm}}|@{\hspace{1mm}}c@{\hspace{1mm}}|c|}
\hline
{\bf Name} &   { $|\mathbf{V}|$} & { $|\mathbf{E}|$ } &  {\bf Weighted?} & {\bf Directed?} & {\bf Size} \\  \hline
USRN &  $24.9$M & $28.9$M  & yes & no & 0.86GB \\ \hline
FB & $58.8$M & $92.2$M  & no & no & 2.42GB \\ \hline
u-BTC & $16.3$M & $95.7$M  & no & no & 1.79GB \\ \hline
u-UKWeb & $6.9$M & $56.5$M  & yes & no & 1.02GB \\ \hline
BTC & $16.3$M & $99.4$M  & no & yes & 1.98GB \\ \hline
Meme & $53.6$M & $117.9$M  & no & yes & 3.17GB \\ \hline
UKWeb & $104$M & $3708$M  & no & yes & 61.8GB \\ \hline
\end{tabular}
\tblcapdown
\end{small}
\end{table}

\subsection{Datasets} \label{sec:exp-data}

We use five real graph datasets as follows: (i) USRN \cite{USRN}, which represents the road network in the US; (ii) FB \cite{gkbm10}, a subgraph of the Facebook friendship graph; (iii) BTC \cite{BTC}, a semantic graph; (iv) Meme \cite{lbkk09} and UKWeb \cite{UKWeb}, which are two web graphs. Among them, only USRN and FB are undirected. Since VC-Index, EM-BFS, and EM-Dijk are all designed for undirected graphs only, we are indeed of more undirected datasets for experiments. For this purpose, we transform BTC and UKWeb into undirected graphs, using the same approach as in previous work (see \cite{ckcc12} for details). After that, for each undirected (resp.\ directed) graph $G$, we compute its largest connected component (resp.\ weakly connected component) $C$, and we use $C$ for experiments. Table~\ref{tbl:exp-data} illustrates the details of the largest component obtained from each graph. In particular, u-BTC and u-UKWeb are obtained from the undirected versions of BTC and UKWeb, respectively.





\header
{\bf Remark.} The previous experimental study on VC-Index \cite{ckcc12} uses USRN, u-BTC, and u-UKWeb instead of their largest connected components (CC) for experiments. We do not follow this approach as it leads to less meaningful results. To explain, consider a massive undirected graph $G$ where each CC is small enough to fit in the main-memory. On such a graph, even if a disk-based method can efficiently answer SSD queries, it does not necessarily mean that it is more scalable than a main-memory algorithm. In particular, one can easily answer an SSD query from any node $s$ in $G$, by first reading into memory the CC that contains $s$, and then running a main-memory SSD algorithm on the CC. In general, given any graph $G$, one can use an I/O efficient algorithm \cite{mr99} to pre-compute the (weakly) connected components in $G$, and then handle SSD queries on each component separately.
%
%

\subsection{Results on Undirected Graphs} \label{sec:exp-undirected}

In the first sets of experiments, we evaluate the performance of each method on four undirected graphs: USRN, FB, u-BTC, and u-UKWeb. For HoD, EM-BFS, and EM-Dijk, we limit the amount of memory available to them to $1$GB, which is smaller than the sizes of all datasets. For VC-Index, we test it with $2$GB memory as it cannot handle any of our datasets under a smaller memory size.

Table~\ref{tbl:exp-pre} shows the preprocessing time of HoD and VC-Index on the four graphs. (EM-BFS and EM-Dijk are omitted as they do not require any pre-computation.) In all cases, HoD incurs a significantly smaller overhead than VC-Index does. In particular, on FB, the preprocessing time of HoD is more than ten times smaller than that of VC-Index. Table~\ref{tbl:exp-space} compares the space consumptions of HoD and VC-Index. Except for the case of u-BTC, the space required by VC-Index is consistently larger than that by HoD.

\begin{table}[t]
\centering
\vspace{-2mm}
\begin{small}
\caption{Preprocessing time (in minutes).}
\label{tbl:exp-pre}
\tblup
\begin{tabular}{|c|c|c|c|c|}
\hline
{\bf Method} &   {\bf USRN} & {\bf FB} &  {\bf u-BTC} & {\bf u-UKWeb} \\  \hline
HoD & $\;\,$4.0 & $\;\,$22.4 & 34.4 & 105.5 \\ \hline
VC-Index & 20.3 & 281.8 & 78.1 & 768.2 \\ \hline
\end{tabular}
\tblcapdown
\end{small}
\end{table}

\begin{table}[t]
\centering
\begin{small}
\caption{Space Consumption (in GB).}
\label{tbl:exp-space}
\tblup
\begin{tabular}{|c|c|c|c|c|}
\hline
{\bf Method} &   {\bf USRN} & {\bf FB} &  {\bf u-BTC} & {\bf u-UKWeb} \\  \hline
HoD & 2.5 & 5.1 & 3.8 & $\;\,$3.3 \\ \hline
VC-Index & 4.3 & 8.3 & 1.2 & 14.0 \\ \hline
\end{tabular}
\tblcapdown
\end{small}
\end{table}

\begin{table}[!t]
\centering
\begin{small}
\caption{Average running time for SSD queries (in seconds).}
\label{tbl:exp-ssd}
\tblup
\begin{tabular}{|c|c|c|c|c|}
\hline
{\bf Method} &   {\bf USRN} & {\bf FB} &  {\bf u-BTC} & {\bf u-UKWeb} \\  \hline
HoD & $\;\,\;\,$1.8 & $\;\,\;\,\;\,$3.2 & $\;\,\;\,$1.6 & $\;\,\;\,$1.4 \\ \hline
VC-Index & $\;\,$27.2 & $\;\,\;\,$94.9 & $\;\,$10.1 & $\;\,$70.0 \\ \hline
EM-BFS & $-$ & $\;\,$465.3 & 395.4 & $-$ \\ \hline
EM-Dijk & 430.7 & 1597.4 & 844.1 & 553.8 \\ \hline
\end{tabular}
\tblcapdown
\end{small}
\end{table}

To evaluate the query efficiency of each method, we generate $100$ SSD queries for each dataset, such that the source node of each query is randomly selected. Table~\ref{tbl:exp-ssd} shows the average running time of each approach in answering an SSD query. The query time of HoD is at least an order of magnitude smaller than that of VC-Index. Meanwhile, VC-Index always outperforms EM-BFS, which is consistent with the experimental results reported in previous work \cite{ckcc12}. We omit EM-BFS on USRN and u-UKWeb, since those two graphs are weighted, for which EM-BFS cannot be used to answer SSD queries. Finally, EM-Dijk incurs a larger query overhead than all other methods.

In the next experiment, we demonstrate an application of HoD for efficient graph analysis. In particular, we consider the task of approximating the {\em closeness} for all nodes in a graph $G$, using the algorithm by Eppstein and Wang \cite{ew04}. The algorithm requires executing $k = \ln n/\epsilon^2$ SSD queries from randomly selected source nodes, where $n$ is the number of nodes in $G$ and $\epsilon$ is a parameter that controls the approximation error. Following previous work \cite{ckcc12}, we set $\epsilon = 0.1$.

Table~\ref{tbl:exp-close} shows an estimation of the time required by each method to complete the approximation task. Specifically, we estimate the total processing time of each method as (i) its query time in Table~\ref{tbl:exp-ssd} multiplied by $k$, plus (ii) its preprocessing time (if any). Observe that both EM-BFS and EM-Dijk incur prohibitive overheads -- they require more than a week to finish the approximation task. In contrast, HoD takes at most $2.4$ hours to complete the task, despite that it pays an initial cost for pre-computation. Meanwhile, VC-Index is significantly outperformed by HoD, and it needs around two days to accomplish the task on FB and u-UKWeb.


\begin{table}[t]
\centering
\vspace{-2mm}
\begin{small}
\caption{Estimated time for {\em closeness} computation (in hours).}
\label{tbl:exp-close}
\tblup
\begin{tabular}{|c|c|c|c|c|}
\hline
{\bf Method} &   {\bf USRN} & {\bf FB} &  {\bf u-BTC} & {\bf u-UKWeb} \\  \hline
HoD & $\;\,\;\,$0.9 & $\;\,\;\,$2.0 & $\;\,\;\,$1.3 & $\;\,\;\,$2.4 \\ \hline
VC-Index & $\;\,$13.2 & $\;\,$51.8 & $\;\,\;\,$6.1 & $\;\,$43.1 \\ \hline
EM-BFS & $-$ & 231.1 & 182.2 & $-$ \\ \hline
EM-Dijk & 203.2 & 793.3 & 389.0 & 240.0 \\ \hline
\end{tabular}
\tblcapdown
\end{small}
\end{table}


\begin{table} [t]
\centering
\begin{small}
\caption{Performance of HoD on directed graphs.}
\label{tbl:exp-directed}
\tblup
\begin{tabular}{| c | c | c | c | } \hline
{\bf Dataset} & {\bf Preprocessing} & {\bf Index Size} & {\bf SSD Query Time} \\ \hline
BTC & 11.4 minutes & $\;\,$2.1 GB & $\;\,$2.6 sec \\ \hline
Meme & $\;\,$1.2 minutes & $\;\,$2.3 GB & $\;\,$1.8 sec\\ \hline
UKWeb & 9.2 hours$\;\,$ & 72.6 GB & 53.7 sec \\ \hline
\end{tabular}
\tblcapdown
\end{small}
\end{table}

\subsection{Results on Directed Graphs} \label{sec:exp-directed}

Our last experiments focus on the three directed graphs: BTC, Meme, and UKWeb. We run HoD on BTC and Meme with 1GB memory, and on UKWeb with 16GB memory, as the enormous size of UKWeb leads to a higher memory requirement. Table~\ref{tbl:exp-directed} shows the preprocessing and space overheads of HoD, as well as its average query time in answering $100$ randomly generated SSD queries on each dataset. (We omit VC-Index, EM-BFS, and EM-Dijk as they do not support directed graphs.) On BTC and Meme, HoD only incurs small pre-computation costs and moderate space consumptions. On UKWeb, the preprocessing, space, and query overheads of HoD are considerably higher, but are still reasonable given that UKWeb contains $30$ times more edges than BTC and Meme do. To our knowledge, this is the first result in the literature that demonstrates practical support for SSD queries on a billion-edge graph. 

\section{Related Work} \label{sec:related}

As mentioned in Section~\ref{sec:intro}, the existing techniques for I/O-efficient SSD and SSSP queries include VC-Index \cite{ckcc12} and a few methods that adopt Dijkstra's algorithm \cite{mo09,mz03,mz12,m09,ks96}. All of those techniques are exclusively designed for undirected graphs, and they incur significant query overheads, as is shown in our experiments. In contrast, HoD supports both directed and undirected graphs, and it offers high query efficiency along with small costs of pre-computation and space.


Other than the aforementioned work, there exists large body of literature on in-memory algorithms for shortest path and distance queries (see \cite{s12,dss09,zmxl13,tsp11} for surveys). The majority of those algorithms focus on two types of queries: (i) {\em point-to-point shortest path (PPSP)} queries, which ask for the shortest path from one node to another, and (ii) {\em point-to-point distance (PPD)} queries, which ask for the length of the shortest path between two given nodes. These two types of queries are closely related to SSD and SSSP queries, in the sense that any SSD (resp.\ SSSP) query can be answered using the results of $n$ PPD (resp.\ PPSP) queries, where $n$ is the number of nodes in the graph. Therefore, it is possible to adopt a solution for PPD (resp.\ PPSP) queries to handle SSD (resp.\ SSSP) queries. Such an adoption, however, incurs significant overheads, especially when $n$ is large. For example, the state-of-the-art solution \cite{adgw11} for PPD queries requires $266$ns to answer a PPD query on the USRN dataset in Section~\ref{sec:exp}. (Note: the solution is not I/O efficient and it requires $25.4$GB memory to handle USRN.) If we use this solution to answer an SSD query on USRN, then we need to execute $24.5$ million PPD queries, which takes roughly $266\textrm{ns} \times 24.5 \times 10^6 = 6.52\textrm{s}$. In contrast, HoD requires only $1.8$s to process an SSD query on USRN, using only $1$GB memory.

Furthermore, almost all existing solutions for PPD and PPSP queries require that the dataset fits in the main memory during pre-computation and/or query processing. This renders them inapplicable for the massive disk-resident graphs considered in this paper. The only exception that we are aware of is a concurrent work by Fu et al.\ \cite{fwcc13}, who propose an I/O-efficient method called {\em IS-Label}. HoD and IS-Label's preprocessing algorithms are similar in spirit, but their index structures and query algorithms are drastically different, as they are designed for different types of queries. In particular, IS-Label focuses on PPD and PPSP queries, and does not efficiently support SSD or SSSP queries.

Finally, we note that previous work \cite{ss05,gss08,dss09} has exploited the idea of augmenting graphs with shortcuts to accelerate PPD and PPSP queries. Our adoption of shortcuts is inspired by previous work \cite{ss05,gss08,dss09}, but it is rather non-trivial due to the facts that (i) we address I/O efficiency under memory-constrained environments, while previous work \cite{ss05,gss08,dss09} focuses on main memory algorithms; (ii) we tackle SSD and SSSP queries instead of PPD and PPSP queries; (iii) we focus on general graphs, while pervious work \cite{ss05,gss08,dss09} considers only road networks (where each node is degree-bounded).

\section{Conclusions}\label{sec:conclu}


This paper presents HoD, a practically efficient index structure for distance queries on massive disk-resident graphs. In particular, HoD supports both directed and undirected graphs, and it efficiently handles {\em single-source shortest path (SSSP) queries} and {\em single-source distance (SSD) queries} under memory-constrained environments. This contrasts the existing methods, which either (i) require that the dataset fits in the main memory during pre-computation and/or query processing, or (ii) support only undirected graphs. With extensive experiments on a variety of real-world graphs, we demonstrate that HoD significantly outperforms the state of the art in terms of query efficiency, space consumption, and pre-computation costs. For future work, we plan to investigate how HoD can be extended to (i) support point-to-point shortest path and distance queries and (ii) handle dynamic graphs that change with time.

\balance

\bibliographystyle{abbrv}
\bibliography{ref}

\appendix

\section{Proof of Theorem 1} \label{sec:proofs}

Our proof of Theorem~\ref{thrm:sssp-correct} utilizes the concepts of {\em rank sequences} and {\em arch paths}, defined as follows.
\begin{definition}[Rank Sequence and Arch Path] \label{def:proof-rank}
Let $\P$ be a path in the augmented graph $\G$ of HoD, such that $\P$ contains $k$ nodes. The {\bf rank sequence} of $\P$ is a sequence of integers $\langle r_1, r_2, \ldots, r_k\rangle$, such that $r_i$ ($i \in [1, k]$) equals the rank of the $i$-th node in $\P$. $\P$ is an {\bf arch path}, if its rank sequence can be divided into three subsequences $\langle r_1, \ldots, r_x\rangle$, $\langle r_x, \ldots, r_y\rangle$, and $\langle r_y, \ldots, r_k\rangle$, such that
\begin{enumerate} [itemsep = 0.5mm, topsep = 2mm]
\item $r_1 < r_2 < \ldots < r_x$, and

\item $r_x = r_{x+1} = \ldots = r_y$, and

\item $r_y > r_{y+1} > \ldots > r_k$. \done
\end{enumerate}
\end{definition}
For instance, consider the path $\P = \langle v_1, v_9, v_{10}, v_8, v_4\rangle$ in Figure~\ref{fig:over-steps}e. The path's rank sequence is $\langle 1, 4, 4, 3, 2\rangle$. This rank sequence can be divided into three subsequences $\langle 1, 4\rangle$, $\langle 4, 4\rangle$, and $\langle 4, 3, 2\rangle$. By Definition~\ref{def:proof-rank}, $\P$ is an arch path.

Let $s$ and $t$ be any two nodes in the original graph $G$, and $P$ be the shortest path from $s$ to $t$ in $G$. If there exist multiple shortest paths from $s$ to $t$, we choose $P$ to be a path where the highest-rank node ranks no lower than any other node on any other shortest path from $s$ to $t$. In the following, we will prove three propositions:
\begin{itemize}
\item {\bf Proposition 1:} For any path $\P$ from $s$ to $t$ in the augmented graph $\G$, its length is no shorter than $P$'s.

\item {\bf Proposition 2:} There exists an arch path $\PA$ in $\G$, such that $l(\PA) = l(P)$, i.e., $\PA$ and $P$ have the same length.

\item {\bf Proposition 3:} When HoD's answers an SSD query from $s$, it will traverse a path no longer than $\PA$.
\end{itemize}
The combination of the above three propositions will establish Theorem~\ref{thrm:sssp-correct}.

\header
{\bf Proof of Proposition 1.} If all edges in $\P$ appear in the original graph $G$, the proposition trivially holds since any path from $s$ to $t$ in $G$ should be no shorter than $P$. In the following, we consider only the case when $P'$ contains at least one shortcut, and we will prove that there exists a path $P'$ in $G$, such that $l(\P) = l(P') \ge l(P)$.

Assume without loss of generality that $\P$ consists of a sequence of $k$ nodes $\langle v_1, v_2, \ldots, v_k\rangle$, where $v_1 = s$ and $v_k = t$. Further assume that $\langle v_i, v_{i+1}\rangle$ ($i \in [1, k]$) is a shortcut. By the preprocessing algorithm of HoD, this shortcut must be constructed when HoD removes a certain node $v$ from the reduced graph, such that $v_i$ and $v_{i+1}$ are incoming and outgoing neighbors of $v$, respectively. This indicates that the augmented graph $\G$ must contain two edges $\langle v_i, v \rangle$ and $\langle v, v_{i+1} \rangle$, such that their total length equals $l(\langle v_i, v_{i+1}\rangle)$. Given those two edges, we change $\P$ by replacing $\langle v_i, v_{i+1}\rangle$ with a two-hop path $\langle v_i, v, v_{i+1}\rangle$. This results in a modified path from $s$ to $t$ in $\G$ that has the same length with the original $\P$.

If the modified $\P$ does not contain any shortcut, then it is a path in $G$, and hence, $l(\P) \ge l(P)$. On the other hand, if the modified $\P$ contains any shortcut, then we can replace the shortcut with a two-hop path in $\G$, as in the previous case for the shortcut $\langle v_i, v_{i+1}\rangle$. By recursively applying this replacement procedure on the shortcuts in $\P$, we can transform $\P$ into a path $P'$ that (i) goes from $s$ to $t$, (ii) contains only edges in $G$, and (iii) has the same length with $\P$. Given that $P$ is the shortest path from $s$ to $t$ in $G$, we have $l(P) \le l(P') = l(\P)$, which completes the proof.

\header
{\bf Proof of Proposition 2.} The proposition trivially holds if $P$ itself is an arch path in $\G$. In the following, we assume that $P$ is not an arch path, and we show that $P$ can be transformed into a arch path $\PA$ in $\G$ with the same length.

Let $r_{max}$ be the highest rank of the nodes in $P$. Let $v_x$ and $v_y$ be the first and last nodes in $P$ whose ranks equal $r_{max}$. We divide $P$ into three subsequences $P_1$, $P_2$, and $P_3$, such that
\begin{enumerate}
\item $P_1$ is the sequence of nodes in $P$ before $v_x$ (including $v_x$).
\item $P_2$ is the sequence of nodes in $P$ between $v_x$ and $v_y$ (including $v_x$ and $v_y$).
\item $P_3$ is the sequence of nodes in $P$ after $v_y$ (including $v_y$).
\end{enumerate}

Let us first consider $P_1$. We say that a node $v$ in $P_1$ is {\em pit}, if $v$ ranks no higher than the node that immediately precedes $v$ in $P_1$. Among the pits in $P_1$, let $v'$ be one with the lowest rank. Let $u$ (resp.\ $w$) be the node that immediately precedes (resp.\ follows) $v'$ in $P_1$. By the preprocessing algorithm of HoD, when HoD removes $v'$ from the reduce graph, it would generate a candidate edge $e_1 = \langle u, w\rangle$, such that the edge has the same length with the two-hop path $\langle u, v', w\rangle$. If $e_1$ is in $\G$ (i.e., it is retained by HoD during preprocessing), then we transform $P_1$ into another path in $\G$, by using $e_1$ to replace the two-hop path $\langle u, v', w\rangle$ in $P_1$. The resulting path has the same length with $P_1$, and it has one less pits than $P_1$. 

On the other hand, if $e_1$ is not in $\G$, there are two possibilities:
\begin{enumerate} 
\item There already exists an edge $e_2 = \langle u, w\rangle$ in the reduced graph, such that $l(e_2) = l(e_1)$. (Note that $l(e_2) < l(e_1)$ is impossible given Proposition 1 and the fact that $P_1$ is the shortest path from $s$ to $v_x$ in $G$.) In that case, $e_2$ must appear in $\G$. Therefore, if we modify $P_1$ by using $e_2$ to replace the two-hop path $\langle u, v', w\rangle$ in $P_1$, we can still obtain a modified path that retains the length of $P_1$ but contains one less pits.
    
\item There exists a two-hop path $\langle u, v^\diamond, w\rangle$ in the reduced graph, such that (i) $l(e_3) = l(e_1)$, and (ii) $v^\diamond$ has a higher rank than $v'$. (Note that $l(e_3) < l(e_1)$ cannot occur due to Proposition 1 and the fact that $P_1$ is the shortest path from $s$ to $v_x$ in $G$.) In that case, we transform $P_1$ by replacing $v'$ with $v^\diamond$. This may not decrease the number of pits in $P_1$, but it retains the length of $P_1$ and replaces a node in $P_1$ with a higher rank node. 
\end{enumerate}
In summary, the above transformation procedure preserves the length of $P_1$, and it either (i) reduces the number of pits in $P_1$ or (ii) substitute a node in $P_1$ with a higher-rank node. Given that the ranks of nodes are bounded, if we recursively apply the procedure on the lowest-rank pit in $P_1$, eventually we should obtain a path $P_1^*$ in $\G$ without any pit, such that (i) $l(P_1^*) = l(P_1)$, and (ii) $P_1^*$ starts at $s$ and ends at $v_x$. In that case, the rank sequence of $P_1^*$ must be an ascending sequence.

Now consider $P_2$ and $P_3$. We say that a node in $P_2$ is a pit if its rank is smaller than $r_{max}$, and we define a node $v$ in $P_3$ as a pit if $v$ ranks no higher than the node that immediately follows $v$ in $P_3$. By applying the same transformation procedure as in the case of $P_1$, we can convert $P_2$ and $P_3$ into two paths $P_2^*$ and $P_3^*$ in $\G$, such that (i) $P_2^*$ and $P_3^*$ have no pit, (ii) $l(P_2^*) = l(P_2)$ and $l(P_3^*) = l(P_3)$, (iii) $P_2^*$ starts at $v_x$ and ends at $v_y$, and (iii) $P_3^*$ starts at $v_y$ and ends at $t$. It can be verified that all nodes in $P_2^*$ should have a rank $r_{max}$, and $P_3^*$'s rank sequence should be a descending sequence.

Let $\PA$ be a path in $\G$ obtained by concatenating $P_1^*$, $P_2^*$, and $P_3^*$. By Definition~\ref{def:proof-rank}, $\PA$ is an arch path, which completes the proof.

\header
{\bf Proof of Proposition 3.} Let $v_x$, $v_y$, $\PA$, $P_1^*$, $P_2^*$, and $P_3^*$ be as defined in the proof of Proposition 2. Without loss of generality, assume that each of $P_1^*$, $P_2^*$, and $P_3^*$ contains at least two nodes. We will prove the proposition by showing that, given an SSD query from $s$, the query processing algorithm of HoD will traverse (i) a path from $s$ to $v_x$ that is no longer than $P_1^*$, (ii) a path from $v_x$ to $v_y$ that is no longer than $P_2^*$, and (iii) a path from $v_y$ to $t$ that is no longer than $P_3^*$.

Recall that HoD's query algorithm consists of three phases: a forward search in the forward graph $\GF$, followed by a core search in the core graph $\GC$, and finally a backward search in the backward graph $\GB$. The forward search is a variant of Dijkstra's algorithm that follows only the outgoing edges whose endpoints rank higher than the starting points. Since the rank sequence of $P_1^*$ is in ascending order, $P_1^*$ should be in the search space of the forward search. Furthermore, by Proposition 1 and the construction of $P_1^*$, $\GF$ does not contain any path from $s$ to $v_x$ that is shorter than $P_1^*$. Therefore, when the forward search terminates, HoD should either identify $P_1^*$ as the shortest path from $s$ to $v_x$ in $\GF$, or identify another path from $s$ to $v_x$ that is no longer than $P_1^*$. In either case, HoD will correctly derive $dist(s, v_x)$, i.e., the distance from $s$ to $v_x$ in $G$.

Now consider $P_2^*$, where each node has the same rank. By the preprocessing algorithm of HoD, all nodes in $P_2^*$ must be in the core graph $\GC$, since any node not in the core graph can only have outgoing edges to higher-rank nodes (see Section~\ref{sec:pre}). Meanwhile, recall that the core search of HoD is a continuation of the forward search in $\GC$, using Dijkstra's algorithm. By the correctness of Dijkstra's algorithm and the fact that $P_2^*$ is the shortest path from $v_x$ to $v_y$ in $\GC$, the core search of HoD should traverse a path from $v_x$ to $v_y$ that is no longer than $P_2^*$.

It remains to prove that HoD's backward search will traverse $P_3^*$. Assume without loss of generality that $P_3^*$ contains a sequence of $k$ nodes $\langle v_1, v_2, \ldots, v_k$, where $v_1 = v_x$ and $v_k = t$. Recall that the backward search of HoD examines nodes in descending order of their ranks. Since $P_3^*$ has a descending rank sequence, the backward search of HoD should examine $v_i$ before $v_{i+1}$, for any $i \in [1, k-1]$.

We will prove by induction that, for any $v_i$ ($i \in [1, k]$), the backward search can correctly derive $dist(s, v_i)$. First, given that $P_1^*$ and $P_2^*$ have been traversed by HoD before the backward search, HoD should be able to compute the precise value of $\d(s, v_1)$ when after it visits $v_1$. Now assume that, after examining $v_j$ ($j \in [1, k-1]$), HoD correctly calculates $\d(s, v_j)$. Then, when HoD inspects $v_{j+1}$, it would identify $dist(s, v_j) + l(\langle v_j, v_{j+1}\rangle)$ as the length of a path from $s$ to $v_{j+1}$. Given that $P_1^*$ is a shortest path from $s$ to $t$ in $\G$ and that $v_{j+1}$ immediately follows $v_j$ on $P_1^*$, we have $dist(s, v_{j+1}) = dist(s, v_j) + l(\langle v_j, v_{j+1}\rangle)$. This indicates that HoD will correctly derive $dist(s, v_{j+1})$, which completes the proof.

\end{sloppy}
\end{document}